# Orbital ordering and fluctuations in a kagome superconductor CsV$_3$Sb$_5$

D. W. Song[1‡], L. X. Zheng[1‡], F. H. Yu[1], J. Li[1], L. P. Nie[1], M. Shan[1], D. Zhao[1], S. J. Li[1], B. L. Kang[1], Z. M. Wu[1], Y. B. Zhou[1], K. L. Sun[1], K. Liu[1], X. G. Luo[1,2], Z. Y. Wang[2], J. J. Ying[2], X. G. Wan[3,6], T. Wu[1,2,4,6†] and X. H. Chen[1,2,4,5,6*]

1. Hefei National Laboratory for Physical Sciences at the Microscale, University of Science and Technology of China, Hefei, Anhui 230026, China

2. CAS Key Laboratory of Strongly-coupled Quantum Matter Physics, Department of Physics, University of Science and Technology of China, Hefei, Anhui 230026, China

3. National Laboratory of Solid State Microstructures and School of Physics, Nanjing University, Nanjing 210093, China

4. CAS Center for Excellence in Superconducting Electronics (CENSE), Shanghai 200050, China

5. CAS Center for Excellence in Quantum Information and Quantum Physics, Hefei, Anhui 230026, China

6. Collaborative Innovation Center of Advanced Microstructures, Nanjing University, Nanjing 210093, China

‡these authors contributed equally to this work
†wutao@ustc.edu.cn
*chenxh@ustc.edu.cn

**Abstract:**
**Recently, competing electronic instabilities, including superconductivity and density-wave-like order, have been discovered in vanadium-based kagome metals AV$_3$Sb$_5$ (A = K, Rb, Cs) with a nontrivial band topology. This finding stimulates wide interests to study the interplay of these competing electronic orders and possible exotic excitations in the superconducting state. Here, in order to further clarify the nature of density-wave-like transition in these kagome superconductors, we performed $^{51}$V and $^{133}$Cs nuclear magnetic resonance (NMR) measurements on the CsV$_3$Sb$_5$ single crystal. A first-order phase transition associated with orbital ordering is revealed by observing a sudden splitting of orbital shift in $^{51}$V NMR spectrum at the structural transition temperature $T_s$ ~ 94 K. In contrast, the quadrupole splitting from a charge-density-**

**wave (CDW) order on $^{51}$V NMR spectrum only appears gradually below $T_s$ with a typical second-order transition behavior, suggesting that the CDW order is a secondary electronic order. Moreover, combined with $^{133}$Cs NMR spectrum, the present result also confirms a three-dimensional structural modulation with a $2a \times 2a \times 2c$ period. Above $T_s$, the temperature-dependent Knight shift and nuclear spin-lattice relaxation rate ($1/T_1$) further indicate the existence of remarkable magnetic fluctuations from vanadium $3d$ orbitals, which are suppressed due to orbital ordering below $T_s$. The present results strongly support that, besides CDW order, the previously claimed density-wave-like transition also involves a dominant orbital order, suggesting a rich orbital physics in these kagome superconductors.**

In recent decades, the exotic electronic states in kagome lattice which holds a special geometric frustration have stimulated considerable interests and becomes a new frontier in condensed matter physics [1-22]. In contrast to the strong coupling limit at half filling [1,2], the exploration of correlated electronic states in kagome lattice with intermediate coupling is still very limited [4,5,6]. Although some intriguing electronic orders (such as chiral *d*-wave/*f*-wave superconductivity) has been theoretically proposed for kagome Hubbard model [4,5,6], their experimental realization is still elusive in the kagome materials, especially for exotic superconductivity. Very recently, the discovery of superconductivity in vanadium-based kagome metal $AV_3Sb_5$ (A = K, Rb, Cs) with superconducting temperature $T_c \sim 3$ K has successfully attracted numerous attentions in community [23-26], which offers a fertile playground to explore the correlation-driven exotic electronic states in kagome lattice.

Besides superconductivity, a density-wave-like transition has also been observed in these kagome metals with transition temperature ($T_s$) ranging from 78 K to 104 K [23]. Earlier X-ray scattering experiment observed an in-plane translational symmetry breaking with a $2a \times 2a$ period below $T_s$, suggesting a possible charge-density-wave (CDW) order [24]. Subsequently, scanning tunneling microscopy (STM) experiment successfully observed a $2a \times 2a$ charge modulation with a novel chiral anisotropy in $KV_3Sb_5$ [27]. More recent STM and X-ray experiments further idenified a three-dimensional (3D) charge modulation with a $2a \times 2a \times 2c$ period in $CsV_3Sb_5$ [28,29]. Theoretically, density functional theory (DFT) calculation also confirms a 3D CDW state and suggests a Peierls-instability-driven CDW picture [30]. Meanwhile, a so-called chiral flux phase has also been proposed as the possible ground state instead of the $2a \times 2a$ charge order [31], in which time reversal symmetry should be broken. Interestingly, besides the in-plane $2a \times 2a$ charge order, an additional $4a$ unidirectional charge order has also been observed on the surface by STM experiments in CsV$_3$Sb$_5$ [32,33], whose underlying mechanism is still elusive. Furthermore, the interplay of superconductivity with CDW order has also been explored by utilizing high-pressure technique in CsV$_3$Sb$_5$ [34-38]. Distinct from the usual competing phase diagram between superconductivity and CDW order, the pressure-dependent $T_c$ in CsV$_3$Sb$_5$ exhibits a nonmonotonic behavior while the density-wave-like transition is continuously suppressed with increasing pressure below 2 GPa [35], suggesting an unusual interplay between superconductivity and CDW order. On the other hand, besides the above mentioned CDW ordering below $T_s$, a possible orbital order may also emerge in this system below $T_s$. Usually, due to the coupling between the orbital and lattice degrees of freedom, orbital ordering would inevitably lead to a structural phase transition [39]. In these new kagome superconductors AV$_3$Sb$_5$, considering a distorted octahedral crystal field on vanadium (V) sites due to six anionic coordinations atoms (Fig.1(a)) and actual $3d$ electron filling, it is possible to hold nearly degenerate $3d$ orbitals in the ground state after the crystal field splitting, which allows for a possible orbital ordering below $T_s$ [40]. Whether orbital ordering is really involved in the density-wave-like transition deserves further investigation. Here, in order to further clarify the exact nature of density-wave-like transition in these kagome superconductors, we performed $^{51}$V and $^{133}$Cs nuclear magnetic resonance (NMR) measurements on CsV$_3$Sb$_5$ single crystal. Evidences for orbital ordering and orbital fluctuations are unambiguously identified, suggesting a rich orbital physics in these kagome superconductors.

As shown in Fig.1(a), the sublattice consisting of V atoms forms an intriguing two-dimensional (2D) kagome net, in which each V atom has six antimony (Sb) atoms as anionic coordination. The six coordinating Sb atoms can be further divided into two kinds of Sb sites at the high-temperature structural phase, including two in-plane Sb(1) sites and four out-of-plane Sb(2) sites. The cesium (Cs) atoms are on the top of Sb(1) sites and individually form an isolating layer between two neighboring [V$_3$Sb$_5$]$^-$ layers. Cs$^+$ layers and [V$_3$Sb$_5$]$^-$ layers are alternatively stacked along c axis, forming a layered structure with the space group of P6/mmm [23]. In principle, NMR spectra is sensitive to the change of local structural environment around the measured nuclei. Here, we performed NMR measurements on both $^{51}$V and $^{133}$Cs nuclei. The nuclear spin number ($I$) is 7/2 for both $^{51}$V and $^{133}$Cs nuclei. Due to electric quadrupole interaction between nuclei with $I > 1/2$ and electrons, the single NMR spectrum of high-spin nuclei would split into $2I+1$ transition lines separated by a so-called quadrupole frequency ($\upsilon_{\alpha\alpha}$). As shown in Fig. 1(c) and (d), the high-temperature NMR spectrum of $^{51}$V and $^{133}$Cs only show one set of NMR transition lines with $2I+1$ equally separated peaks, which is consistent with only one structural site for both V and Cs sublattices above $T_s$ as shown in Fig.1(a). Below $T_s$, both NMR spectrum of $^{51}$V and $^{133}$Cs split into two sets of NMR transition lines, indicating two distinct structural sites for both V and Cs sublattices. By considering a first-order approximation, the NMR spectrum of $^{51}$V would be only sensitive to the in-plane structural modulation. In contrast, besides in-plane modulation, the NMR spectrum of interlaminar $^{133}$Cs is also dependent on the structural modulation along $c$ axis. As suggested by previous STM and X-ray scattering experiments [28,29], the density-wave-like transition leads to a 3D structural modulation with a $2a\times2a\times2c$ period. In this situation, based on the analysis of NMR spectrum, both NMR spectra of $^{51}$V and $^{133}$Cs should split into two sets of transition lines with equal spectral weight (see details in S.6 section of supplementary materials). This is perfectly consistent with our present observations, further supporting a $2a\times2a\times2c$ period in bulk $CsV_3Sb_5$.

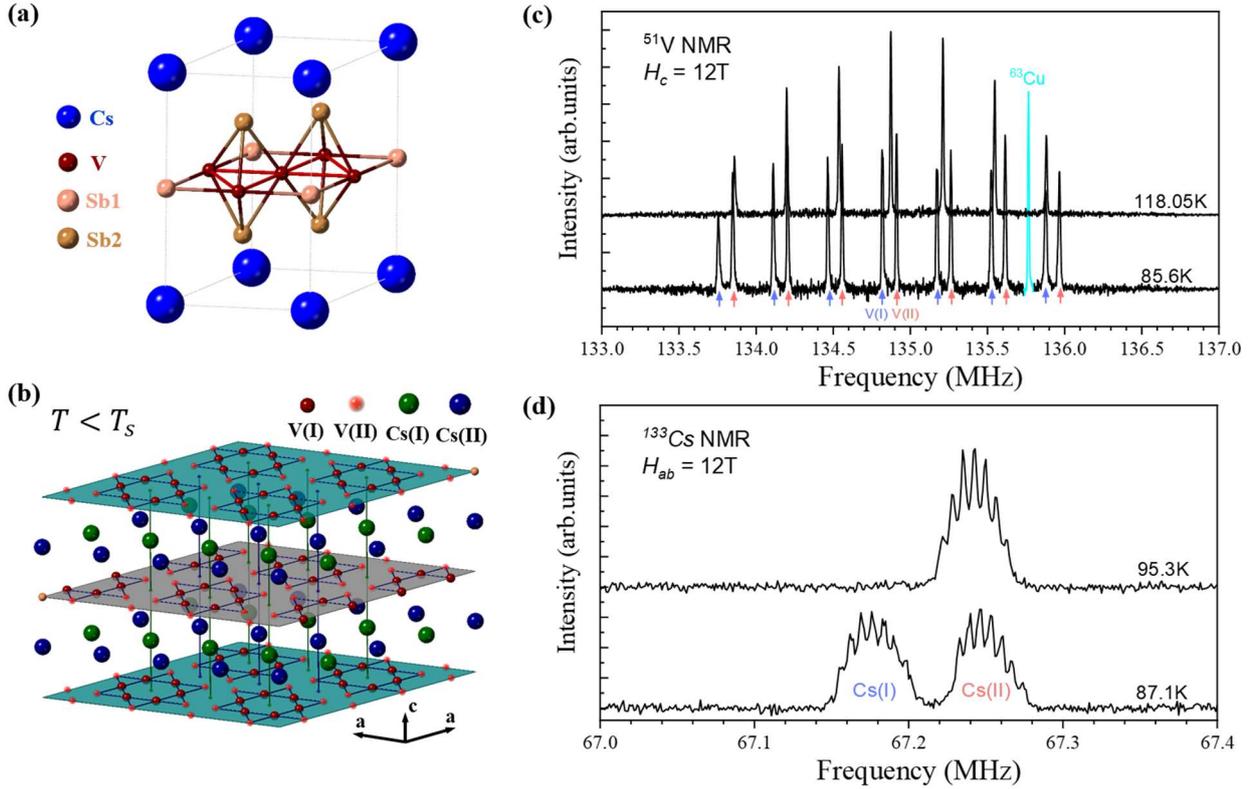

Fig. 1. Evidence for a 3D structural modulation by $^{51}$V and $^{133}$Cs NMR. (a) Crystal structure of CsV$_3$Sb$_5$ above $T_s$; (b) Illustration of the formation of an in-plane Star of David structure and different V/Cs sites under a $2a \times 2a \times 2c$ superstructure below $T_s$; (c) The representative $^{51}$V NMR spectra measured above and below $T_s$ with the external magnetic field $H$ = 12T along $c$ axis ($H_c$); (d) The representative $^{133}$Cs NMR spectra measured above and below $T_s$ with $H$ = 12T parallel to $ab$ plane ($H_{ab}$). The present NMR spectrum indicate that there are two inequivalent V and Cs sites below $T_s$, which are labeled as V(I), V(II), and Cs(I), Cs(II), respectively.

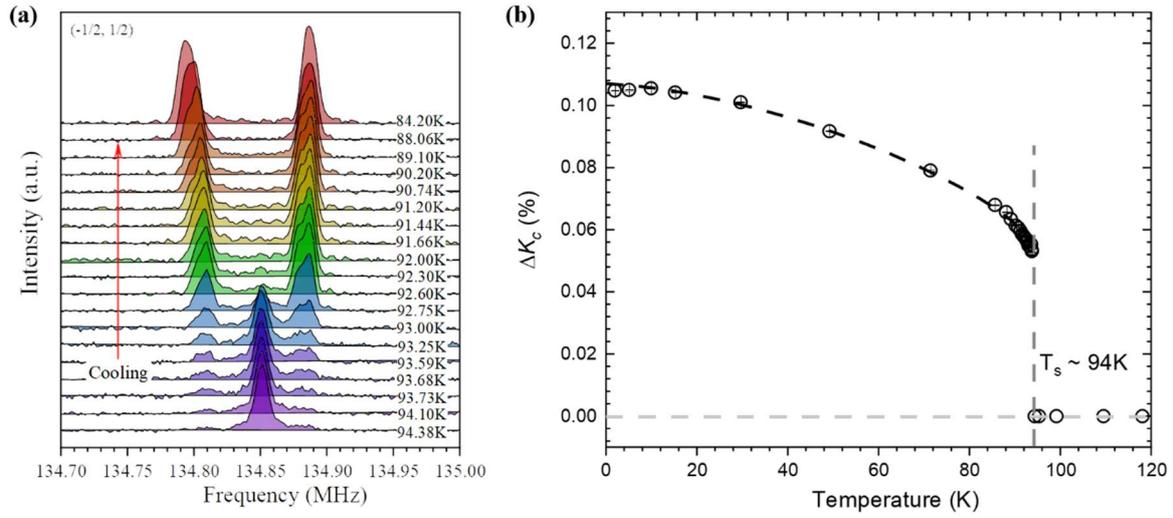

Fig. 2. Evidence for a first-order phase transition associated with orbital ordering by $^{51}$V NMR. (a) The temperature dependence of the central transition lines of $^{51}$V NMR with the temperature cooling across $T_s$. (b) Temperature dependence of the splitting of Knight shift ($\Delta K_c$), which is obtained by calculating the difference of Knight shift between V(I) and V(II) sites. The coexistence of high-temperature and low-temperature phases between 92 K and 94.38 K as well as a sudden jump of $\Delta K_c$ manifest a first-order phase transition associated with an orbital ordering at $T_s$.

Next, we would discuss the origin of the observed NMR splitting at $^{51}$V nuclei. In general, both magnetic and quadrupole interaction between nuclei and electrons can produce the splitting in NMR spectrum below $T_s$. However, they would give different manifestations in NMR spectrum. As shown in Fig.1(c), the splitting on each NMR transition line is almost the same, which indicates that the predominant contribution of splitting is not from quadrupole interaction but magnetic interaction. In order to extract the temperature-dependent behavior of the magnetic splitting, we measured the temperature-dependent splitting on the central transition lines in great details. Interestingly, as shown in Fig. 2(a), the temperature-dependent central transition lines show a typical behavior for a first-order phase transition, which holds a narrow temperature range to exhibit two-phase coexistence. Here, the temperature range for two-phase coexistence is only about 2K. Moreover, the splitting of NMR spectrum shows a sudden jump at $T_s$, which is also consistent with a first-order phase transition. In supplementary materials, we also measure the field dependence of splitting (see details in S.5 section of supplementary materials), suggesting that such magnetic splitting comes from a splitting of Knight shift instead of a spontaneous internal field. Therefore, time reversal symmetry is still preserved below $T_s$. This is also consistent with previous μSR experiment [41]. The temperature-dependent splitting of Knight shift is plotted in Fig.2(b). After a sudden jump at $T_s$, the splitting of Knight shift shows a continuous increase and finally becomes saturated at the low temperature below 20K.

How to understand the splitting of Knight shift? In principle, the total Knight shift has two major contributions, including spin shift and orbital shift. Spin shift is proportional to spin susceptibility ($\chi_{spin}$) and orbital shift is proportional to orbital susceptibility ($\chi_{orb}$). Here, the splitting of Knight shift is mainly from orbital shift (more detailed explanation in the following discussion), suggesting a possible orbital order below $T_s$. Usually, the orbital susceptibility is due to Van Vleck paramagnetism which is proportional to $1/\Delta$ [40], where $\Delta$ is the energy gap between orbital ground state and excited state due to crystal field splitting. The energy scale of $\Delta$ in 3$d$ transition metal can vary in a wide range of energy scale from ~0.1 eV to ~1 eV [40,42,43]. The change of local structural environment could lead to a corresponding change of local $\Delta$ at different sites. Therefore, a slightly splitting of orbital shift is naturally expected for a structural phase transition. However, orbital order is beyond this trivial situation and leads to different orbital configurations at different sites similar as that in manganites [39]. Strictly speaking, whether the observed splitting of orbital shift comes from an orbital order needs more theoretical calculations on the actual crystal field splitting. However, the magnitude of the observed splitting from orbital shift is so large in our situation, close to 1/3 of the total Knight shift, suggesting a considerable change of $\Delta$ or orbital configurations. A weak structural phase transition without orbital order would not be expected to produce such a large change of orbital shift [42,43]. Therefore, we ascribe the large splitting of orbital shift to an orbital ordering at $T_s$. Later, we will show another evidence for orbital order by measuring orbital fluctuations above $T_s$, which together with the observed orbital shift confirms an orbital order.

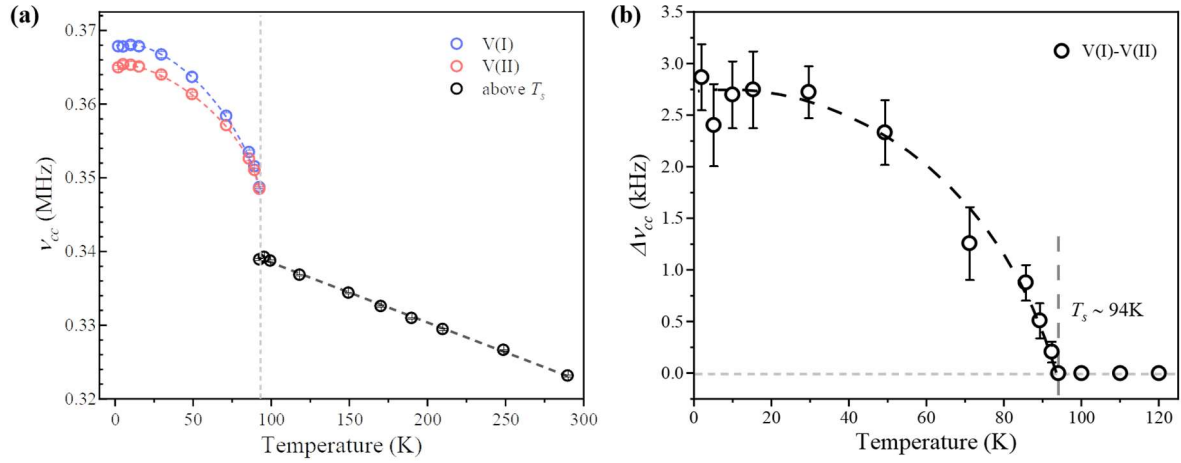

Fig. 3. The quadrupole splitting on $^{51}$V NMR due to a CDW order. (a) Temperature dependence of the quadrupole frequency $\nu_{cc}$, which is extracted from the $^{51}$V NMR spectrum measured with H//c axis. (b) The temperature dependence of $\Delta\nu_{cc}$, which is the difference of $\nu_{cc}$ between V(I) and V(II) sites. In contrast to orbital order, a second-order-like temperature dependence of $\Delta\nu_{cc}$ appears below $T_s$, which is ascribed to a secondary CDW order.

Besides the predominant splitting of Knight shift, there is also a minor splitting from quadrupole interaction on the NMR satellites (more details see the S.1 section of supplementary materials). In principle, the quadrupole frequency is proportional to the local electric field gradient (EFG), which is sensitive to electronic ordering in both orbital and charge degrees of freedom [44,45]. Therefore, both orbital order and CDW order should affect the quadrupole frequency. As shown in Fig.3(a), although the temperature dependent quadrupole frequency ($\nu_{cc}$) shows a sudden jump at $T_s$ similar as that of the splitting of Knight shift, the quadrupole splitting only shows a continuous change below $T_s$ like a second-order phase transition. If considering the accumulating experimental evidences for an in-plane $2a\times 2a$ charge order, the quadrupole splitting should be ascribed to the CDW order which behaves like a secondary electronic order below $T_s$. It is clear that the change of $\nu_{cc}$ is also dominated by orbital order and the contribution from CDW order ($\sim \Delta\nu_{cc}$) is roughly estimated to be only about 10% of the total change of $\nu_{cc}$ below $T_s$. In addition, a $T$-linear behavior is also observed above $T_s$ for the temperature-dependent $\nu_{cc}$. The value of $\nu_{cc}$ is increased by about 5% from room temperature to 100 K, which is one order of magnitude larger than the corresponding change of lattice parameters due to temperature cooling [24]. This implies a possible evolution of orbital states even above $T_s$.

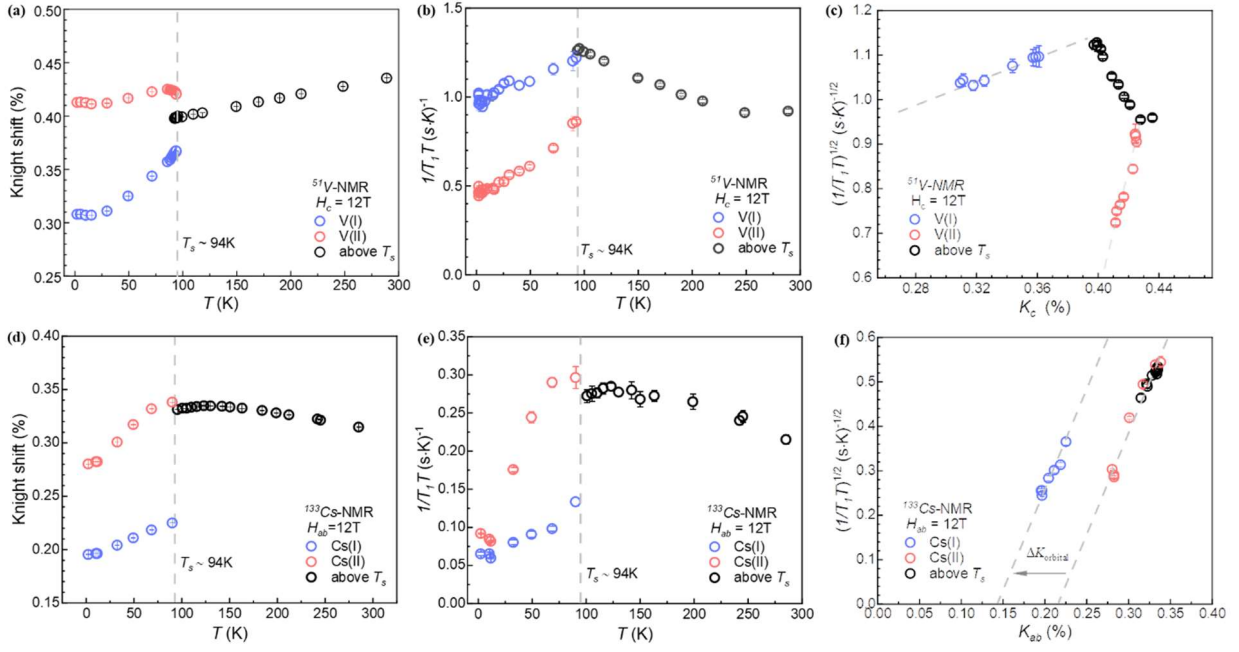

Fig. 4. Temperature-dependent Knight shift and spin-lattice relaxation rate of $^{51}$V and $^{133}$Cs. (a), (b) Temperature dependence of Knight shift and spin-lattice relaxation rate divided by temperature ($1/T_1T$) for $^{51}$V with $H_c$ = 12T; (c) the corresponding Korringa plot by taking data from (a) and (b); (d), (e) Temperature dependence of Knight shift and spin-lattice relaxation rate divided by temperature ($1/T_1T$) for $^{133}$Cs with $H_{ab}$ = 12T; (f) the corresponding Korringa plot by taking data from (d) and (e). All the dash lines are the guides for eyes.

In order to further clarify the orbital physics in CsV$_3$Sb$_5$, we systematically measure the temperature-dependent Knight shift and spin-lattice relaxation rate ($1/T_1$) in a wide temperature range. As shown in Fig. 4(a) and (b), the temperature-dependent Knight shift and $1/T_1T$ of $^{51}$V nuclei show distinct temperature-dependent behaviors above $T_s$. In contrast, as shown in Fig. 4(d) and (e), a similar temperature-dependent behavior of Knight shift and $1/T_1T$ is observed at $^{133}$Cs nuclei. Below $T_s$, besides the splitting of Knight shift, the $1/T_1T$ at different V and Cs sites also show different values. In conventional Fermi-liquid scenario, spin shift ($K_s$) and $1/T_1T$ are both related to the density of states at Fermi level (D($E_F$)), which hold a standard Korringa relation with $1/T_1T \sim K_s^2 \sim $ D($E_F$)$^2$. As shown in Fig. 4(c) and (f), while the Korringa relation remains precisely in the whole temperature range for the

NMR results of $^{133}$Cs nuclei, a clear deviation from Korringa relation is observed for the NMR results of $^{51}$V nuclei. It should be reminded that, if considering a standard Korringa relation between Knight shift and $1/T_1T$, the observed splitting of Knight shift does not follow the expectation from the splitting of $1/T_1T$, which suggests a major contribution of splitting from orbital shift instead of spin shift. Although the splitting of Knight shift at $^{133}$Cs sites also mainly comes from orbital shift, a standard Korringa relation indicates, except the splitting at $T_s$, the temperature dependence of both Knight shift and $1/T_1T$ are dominated by spin susceptibility. In contrast, the temperature-dependence of both Knight shift and $1/T_1T$ at $^{51}$V sites should involve an important contribution from orbital susceptibility. Therefore, it is reasonable to consider that the difference of temperature dependent behavior between $^{51}$V and $^{133}$Cs nuclei is due to a nontrivial contribution from orbital susceptibility. It should be also noted that the temperature dependence of Knight shift at $^{51}$V sites is quite consistent with that of bulk susceptibility above $T_s$ (see the temperature-dependent bulk susceptibility in the S.1 section of supplementary materials), suggesting a predominant orbital contribution also in bulk susceptibility. Usually, the orbital susceptibility is almost temperature independent or increasing with temperature decreasing. The decreasing orbital susceptibility with temperature cooling is quite unusual, suggesting a sophisticated orbital physics. In general, $1/T_1T$ is related to the dynamic spin susceptibility and has a general expression as: $1/T_1T \sim \sum_q A_q^2 \cdot \chi_m''(q,\omega)/\omega$, where $A_q$ is the $q$-dependent hyperfine coupling tensor, $\chi_m''$ is the imaginary part of dynamic magnetic susceptibility and $\omega$ is the Larmor frequency. In fact, $\chi_m''$ can have two different contributions from spin and orbital susceptibility respectively. Here, since a nontrivial orbital contribution only shows up at $^{51}$V site but not at $^{133}$Cs sites, we could have chance to separate spin and orbital contributions in $1/T_1T$ by comparing the temperature dependent behavior at both $^{51}$V and $^{133}$Cs sites. As shown in Fig. 4(b) and (e), although the temperature-dependent $1/T_1T$ at $^{133}$Cs sites shows a saturation below 150 K, the temperature-dependent $1/T_1T$ at $^{51}$V sites still remains an upturn behavior up to $T_s$. This result points out a significant contribution from orbital fluctuations on $1/T_1T$ at $^{51}$V sites. Below $T_s$, the upturn behavior in $1/T_1T$ is strongly suppressed at $^{51}$V sites and both Knight shift and $1/T_1T$ show a similar temperature dependent behavior, indicating the suppression of orbital fluctuations after orbital ordering transition. In addition, besides applying magnetic field along $c$ axis, we also measure the Knight shift and $1/T_1T$ with an in-plane magnetic field above $T_s$ (see details in the S.4 section of supplementary materials). Significant orbital fluctuations above $T_s$ are also confirmed in these results.

So far, our NMR results unambiguously demonstrate a rich orbital physics in CsV$_3$Sb$_5$, including orbital order and relevant orbital fluctuations. How to further understand the exact orbital physics needs more theoretical inputs on these new kagome materials. On the other hand, It would be very interesting to explore the correlation between orbital ordering/fluctuations and superconductivity in this system. Based on the previous high-pressure studies on CsV$_3$Sb$_5$, the superconductivity would be largely enhanced and the transition temperature reaches to a maximum when the density-wave-like transition is completely suppressed around 2.0 GPa [34,35]. Our present NMR results suggest that the orbital fluctuations might be also enhanced around this critical pressure. Whether the optimized superconductivity is related to a possible enhancement of orbital fluctuations deserves further exploration. In addition, a topological charge order with chiral anisotropy has been observed by previous STM experiment [27]. Based on the present NMR results, the observed orbital order and secondary CDW order only breaks translational symmetry but preserves the rotational symmetry. If considering the rotational symmetry breaking at low temperatures as suggested by STM results, another phase transition below $T_s$ would be expected to break the rotational symmetry. More experiments to uncover the possible nematic transition would be urgently desired to clarify this issue.

*Notes: During the preparation of this manuscript, we realized that a similar NMR work on CsV$_3$Sb$_5$ has been posted in arXiv (Chao Mu et al., arXiv:2104.06698). The results of $^{51}$V NMR on the normal state are qualitatively consistent with our present work.*

## Acknowledgments

We thank the valuable discussion with Y. L. Wang, G. Chen and J. F. He. This work is supported by the National Key R&D Program of the MOST of China (Grants No. 2017YFA0303000, 2016YFA0300201), the National Natural Science Foundation of China (Grants No. 11888101, 11522434), the Strategic Priority Research Program of Chinese Academy of Sciences (Grant No. XDB25000000), the Anhui Initiative in Quantum Information Technologies (Grant No. AHY160000).